\documentstyle[preprint,prl,aps]{revtex}

\begin{document}
\draft
\title{Sensitivity to initial conditions and nonextensivity in biological evolution}

\author{Francisco A. Tamarit, Sergio A. Cannas\cite{auth2}}  

\address{Facultad de Matem\'atica, Astronom\'\i a y F\'\i sica, 
Universidad Nacional de C\'ordoba, Ciudad Universitaria, 5000 
C\'ordoba, Argentina}

\author{Constantino Tsallis}

\address{Centro Brasileiro de Pesquisas F\'\i sicas\cite{auth3}, Xavier Sigaud 150,
22290-180 Rio de Janeiro-RJ, Brazil\\
Facultad de Matem\'atica, Astronom\'\i a y F\'\i sica, Universidad Nacional de 
C\'ordoba, Ciudad Universitaria, 5000 C\'ordoba, Argentina\\
Departamento de Fisica, Universidade Federal do Rio Grande do Norte,
59072-970 Natal-RN, Brazil}

\date{\today}
\maketitle

\begin{abstract} 
We consider biological evolution as described within the Bak and Sneppen 1993 
model. We exhibit, at the self-organized critical state, a power-law sensitivity to 
the initial conditions, calculate the associated exponent, and relate it to the recently 
introduced nonextensive thermostatistics. The scenario which here emerges 
{\it without tuning} strongly reminds that of the {\it tuned} onset of chaos in say 
logistic-like onedimensional maps. We also calculate the dynamical exponent z.
\end{abstract}


There is nowadays a massive evidence of fractals and scale invariant phenomena 
in nature. They appear in an impressive variety of inanimate systems such as the 
geological  (e.g., earthquakes) or climatic (e.g., atmospheric turbulence) ones, as 
well as of biological or living systems (e.g., biological evolution, cell growth, 
economic phenomena, among others). In most of the naturally occuring cases, no 
particular tuning is perceived.  Per Bak and collaborators have advanced 
\cite{perbak1,perbak2} the hypothesis that, for many if not all the cases, this is so 
because the microscopic dynamics of the system makes it to spontaneously evolve 
towards a critical, scale-invariant, state. This is largely known today as 
{\it self-organized criticality} (SOC). Models have been formulated and 
experiments have been performed which profusely exhibit this interesting type of 
behaviour in sandpiles, ricepiles, earthquakes and others (see for instance 
\cite{kim1,kim2,kim3} and references therein). One such model is that introduced 
in 1993 by  Bak and Sneppen \cite{perbak3} to paradigmatically describe biological 
species evolution. This is the model that we focuse herein. A variety of its properties 
are already known. Nevertheless, there is a crucial one that has never been 
addressed, namely the {\it sensitivity to the initial conditions}, which is known to be 
most relevant in nonlinear dynamical systems (quantities intensively studied such as 
Liapunov exponents, spread of damage, are in fact nothing but specific expressions 
of this concept). The study of this important property is the basic aim of the present 
work.

Before describing our particular approach of this evolution model, let us introduce 
some preliminary notions by using, as a simple illustration, the following 
one-dimensional logistic-like map (see \cite{hauser} and references therein)
\begin{equation}
x_{t+1} \, = \, 1 \, - \, a \, \left | x_t \right |^\zeta, \,\,\, 
(\zeta > 1; 0<a\le 2; t=0,1,2, \ldots )
\end{equation} 
This map recovers, for $\zeta=2$, the usual logistic map (in its centered 
representation). For fixed $\zeta$ there is a critical value $a_c(\zeta)$ such that , 
for $a<a_c(\zeta)$, we observe a regular evolution (finite-cycle attractors), whereas, 
for $a>a_c(\zeta)$, chaos becomes possible. Approaching $a_c(\zeta)$ from below we 
can see the celebrated doubling-period road to chaos, with its successive 
bifurcations. Topological properties of the evolution (such as the nature of the 
successive attractors while varying $a$) do not depend on $\zeta$, but metrical 
properties (such as the Feigenbaum's constants, characterizing geometrical rates 
of approach of the bifurcations leading to the chaotic critical point $a_c(\zeta)$) 
do depend on $\zeta$. (Let us anticipate that the quantity we shall focuse herein, 
the entropic index $q$, belongs to this second class of properties). If we consider, 
at $t=0$, two values of $x_0$ which slightly differ by $\Delta x(0)$ and follow 
their time evolution, we typically observe the following exponential behaviour for 
$\Delta x(t)$
\begin {equation} 
\lim_{\Delta x(0) \rightarrow 0}\frac{\Delta x(t)}{\Delta x(0)}= 
\exp [\lambda_1\;t] \;\;\;
\end {equation}
If $\lambda_1<0$ (which is in fact the case for most values of $a$ {\it below} 
$a_c(\zeta)$) we shall say that the system is {\it strongly insensitive} to the initial 
conditions. If $\lambda_1>0$ (which is in fact the case for most values of $a$ 
{\it above} $a_c(\zeta)$) we shall say that the system is {\it strongly sensitive} 
to the initial conditions. Finally, if $\lambda_1$ vanishes we shall speak of a 
{\it marginal} case. This is what happens, in particular, for the chaotic critical 
point. For this value of $a$, the sensitivity is not characterized by an 
exponential-law, but rather by a power-law. Eq. (2) can be generalized as 
follows \cite{tpz,maceio}
\begin{equation} 
\lim_{\Delta x(0) \rightarrow 0}\frac{\Delta x(t)}{\Delta x(0)}=
[1+(1-q)\;\lambda_q\;t]^\frac{1}{1-q}\;\;\;
 (q \in {\cal R})
\end{equation}
We immediately see that the $q=1$ case reproduces Eq. (2), whereas 
for $q \ne 0$ we have
\begin{equation} 
\lim_{\Delta x(0) \rightarrow 0}\frac{\Delta x(t)}{\Delta x(0)} \sim
[(1-q)\lambda_q]^{\frac{1}{1-q}}\;t^{\frac{1}{1-q}} 
\,\,\,\, (t \rightarrow \infty)
\end{equation}
The $q<1$ and the $q>1$ cases respectively correspond to {\it weakly sensitive} 
and {\it weakly insensitive} to the initial conditions \cite{tpz}. The coefficient 
$\lambda_q$ is a generalized Liapunov exponent, and satisfies\cite{tpz,maceio} 
$K_q=\lambda_q$ if $\lambda_q \ge 0$ and $K_q=0$ if $\lambda_q <0$ 
(generalization, for arbitrary $q$, of the well known Pesin equality) , where the 
generalized Kolmogorov-Sinai entropy $K_q$ is defined\cite{tpz,maceio} 
analogously to the usual Kolmogorov-Sinai entropy ($K_1$ herein). More 
precisely, in the same way $K_1$ essentially is the increase per unit time of the 
Boltzmann-Gibbs-Shannon entropy $S_1\equiv -\sum_i\,p_i\,ln\,p_i$ , $K_q$ 
essentially is the increase per unit time of the generalized, nonextensive, 
entropic form\cite{tsallis}
\begin{equation}
S_q \equiv \frac{1-\sum_i p_i^q}{q-1}\;\;\;(q \in \cal{R})
\end{equation}
The nonextensivity of this form can be seen from the fact that if $A$ and $B$ 
are two independent systems (in the sense that the probabilities associated with 
$A+B$ factorize into those of $A$ and $B$), then 
\begin{equation}
S_q(A+B)=S_q(A)+S_q(B)+(1-q)\,S_q(A)\,S_q(B) 
\end{equation}
We immediately verify that, since $S_q$ is nonnegative, $q=1$, $q<1$ and 
$q>1$ respectively correspond to the extensive, superextensive and 
subextensive cases.
This generalized entropy (5) has generated a 
generalized thermostatistics\cite{tsallis,curado}(which recovers the usual, 
extensive, Boltzmann-Gibbs statistics as the $q=1$ particular case), and 
has received applications in a variety of situations such as self-gravitating 
systems\cite{plastino,bogho}, two-dimensional-like turbulence in 
pure-electron plasma\cite{bogho,anteneodo}, 
L\'evy-like\cite{levy} and correlated-like\cite{correlated} anomalous 
diffusions, solar neutrino problem\cite{quarati}, peculiar velocity distribution 
of galaxy clusters\cite{lavagno}, cosmology\cite{hamity}, linear response 
theory\cite{rajagopal}, long-range fluid and magnetic systems\cite{n*}, 
optimization techniques\cite{optimization} , among others (including of 
course the nonlinear maps (1)\cite{tpz,maceio}).\\
Now that we have introduced all the needed ingredients, we can close 
the illustration associated with Eq. (1) by mentionning that the 
$\zeta$-dependence of the entropic index $q$ and of the fractal dimension 
$d_f$ have been calculated\cite{maceio} at the chaotic critical point 
$a_c(\zeta)$. It was exhibited that, while $d_f$ varies from 0 to 1 
(or close to it), $q$ increases from  $-\infty$ to 1 (or close to it). 
Therefore, the Boltzmann-Gibbs limit $q=1$ is attained when the attractor 
has an euclidean (nonfractal) dimension $d_f=1$. But, when the system lives 
on a fractal ($d_f<1$ in the present case), nonextensive behaviour is revealed.\\
Let us now return to the Bak and Sneppen evolution model. We shall exhibit 
that, at its self-organized critical state,  weak sensitivity to the initial conditions 
(i.e., a power-law of the type indicated in Eq. (4)) occurs very similarly to the 
one just described for the chaotic critical point of the map (1). The model 
consists in a $N$-site ring (linear chain with periodic boundary conditions); 
on each site ($j=1,2,...,N$) we locate a real variable 
$B_j\,(0 \le B_j \le 1, \,\forall j)$ which corresponds to a "fitness barrier" 
separating two connected (first-neighboring) "species of living organisms". 
We start by randomly and independently attributing the set of values $\{B_j\}$. 
At each successive {\it elementary} time step we identify the {\it smallest} $B_j$, 
and randomly change ("mutate") it as well as its two nearest neighbors. After 
some transient, a peculiar self-organized state emerges\cite{perbak3}, rich in 
avalanches of all sizes and other scale invariant properties which makes the 
system to exhibit a variety of power laws. \\
In the present work we have focused, as follows, the sensitivity to the initial 
conditions of the Bak and Sneppen model. Once SOC has been achieved, we 
consider that system as replica 1 ($\{B_j^{(1)}\}$) and create a replica 2  
($\{B_j^{(2)}\}$) by randomly choosing one of its $N$ sites, and exchanging 
the value associated with this site and that with the smallest barrier; we consider 
this moment as  the {\it collective} time step $t=1$ (we define a collective time 
step as $N$ times the elementary time step, i.e., each site is going to be updated 
only once in average during a unit collective time step). From now on, we apply 
for both replicas the model rules of identifying the smallest barrier and updating 
that particular one and its two first-neighbors. We use (as in usual damage 
spreading techniques) the {\it same random numbers for both replicas} 
(hence, three different and independent random numbers are involved in the 
operation, since we change a particular barrier and its first-neighbors).\\
We define now the {\it Hamming distance} between the two replicas as
\begin{equation}
D(t) \equiv \frac{1}{N}\sum_{j=1}^N\,
\left |B_j^{(1)}(t)\,-\,B_j^{(2)}(t)\right |
\end{equation}
One such realization is shown in Fig. 1 for $N=1000$. We then average 
$N_r$ realizations (we have used typically $N_r\,N\,=\,10^5$) and obtain 
$\langle D \rangle(t)$: see Fig. 2, where a power-law is evident These results 
enable the determination of the slope $1/(1-q)$ (see Eq. (4)), hence of $q$; we 
obtained $q=-2.1$ (to be compared, for instance, with the $\zeta=2$ logistic 
map value 0.24...\cite{tpz}). Finally, for fixed $N$, if we denote by $\tau$ the 
value of $t$ at which the increasing regime {\it crossovers} onto the 
saturation regime (intersection, in Fig. 2, of two straight lines, namely those 
defined by the linearly increasing branch of the curve and the horizontal 
branch. The proportionality $\tau(N) \propto N^z$ defines the dynamical 
exponent $z$ (\cite{dynexp} and references therein). We obtained 
(see Fig.3) $z=1.56$, to be compared, for instance, with 2.16 
obtained\cite{dynexp} for the square-lattice Ising ferromagnet.\\
Since $(q-1)$ measures the degree of entropy nonextensivity (see Eq. (6)), 
it is an important index to be analyzed whenever discussing universality 
classes. Consistently, the determination of $q$ for other SOC models would 
be very welcome. Indeed, it will provide an insight on the fractal nature of 
the attractor towards which the system is spontaneously driven.\\

We are indebted to  A.R. Plastino for many fruitful
suggestions and discussions about this beautiful problem. Two of us
(FAT and SAC) acknowledge warm hospitality received at the Centro 
Brasileiro
de Pesquisas F\'\i sicas (Brazil), where this work was partly
carried out.  This work was partially supported by grants
B11487/4B009 from Funda\c{c}\~ao Vitae (Brazil), PID 3592 from
Consejo Nacional de Investigaciones Cient\'\i ficas y T\'ecnicas
CONICET (Argentina),PID 2844/93 from Consejo Provincial de
Investigaciones Cient\'\i ficas y Tecnol\'ogicas (C\'ordoba,
Argentina) and a grant from Secretar\'\i a de Ciencia y
Tecnolog\'\i a de la Universidad Nacional de C\'ordoba
(Argentina).

\begin{figure}
\caption{Time evolution of the damage associated with one 
realization of a typical system with $N=1000$ .}
\label{fig1}
\end{figure}
\begin{figure}
\caption{Average of $N_r$ realizations such as those of figure 
1 for three different sizes: $N=1000$ (top), $N=500$ (middle) 
and $N=250$ (bottom). The slope $1/(1-q)$ equals 
$0.32 \pm 0.01$.}
\label{fig2}
\end{figure}
\begin{figure}
\caption{Log-log plot of  $\tau$ versus $N$ for the three 
curves of figure 2 .~~~~~~~~~~ ~~~~~~~~~~~~~~~~~~~~}
\label{fig3}
\end{figure}

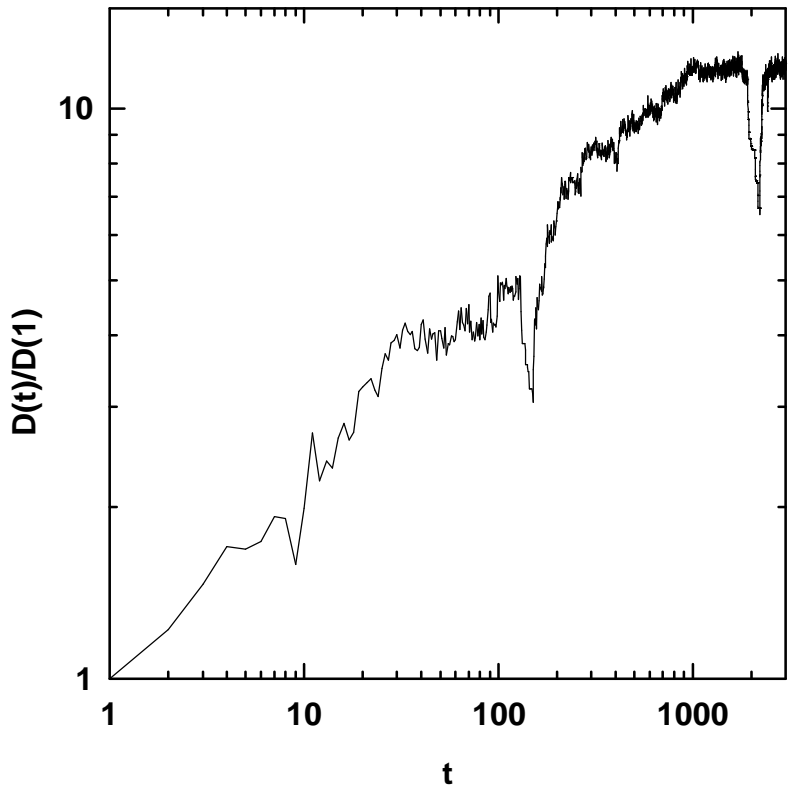
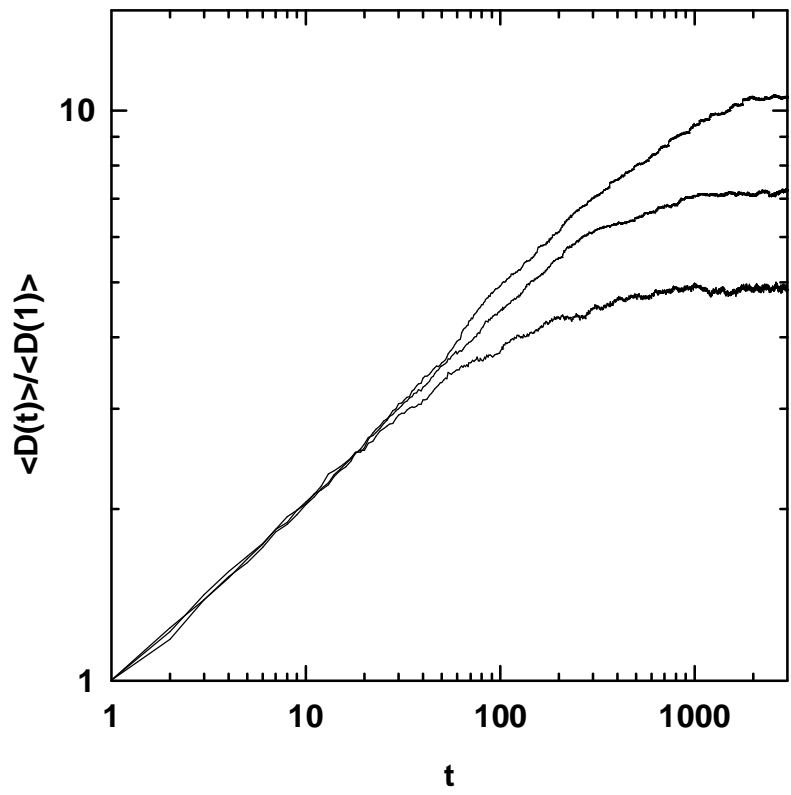
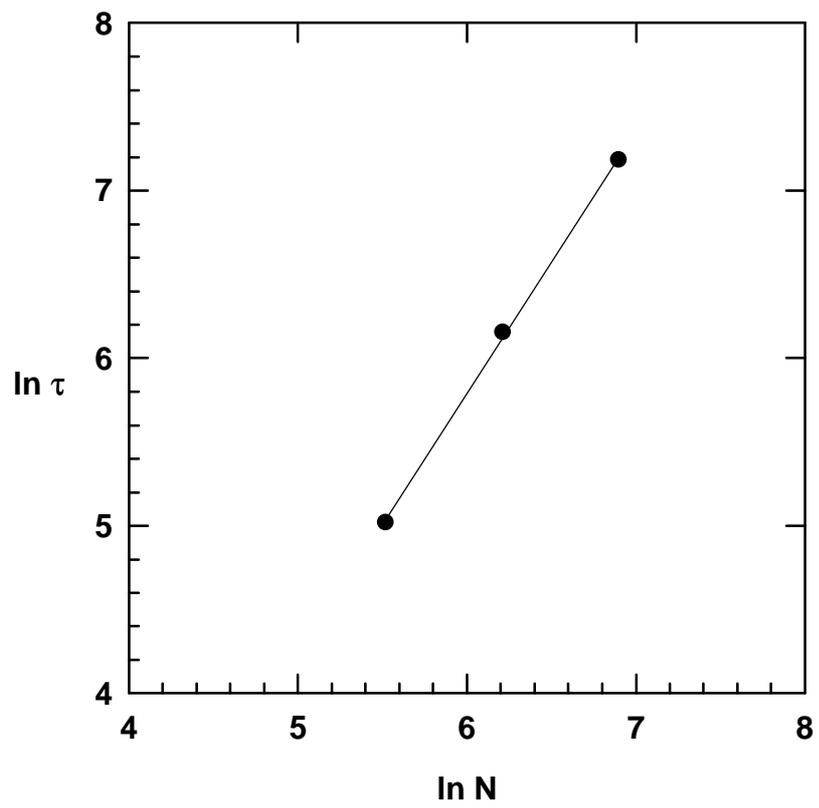
\end{document}